# GEOMETRIC QUANTIZATION BY PATHS
### PART III: THE METAPLECTIC ANOMALY

PATRICK IGLESIAS-ZEMMOUR

ABSTRACT. In the previous parts of this work, we established the Prequantum Groupoid $\mathbf{T}_\omega$ as the universal geometric container for quantum mechanics. This approach, which we call the *Geometric Quantization by Paths* (GQbP) framework, replaces the traditional construction of principal bundles with the distillation of the space of histories. In this third part, we cross the "Threshold of Analysis" by constructing the intrinsic observable algebra of the system. The harmonic oscillator is treated here as a validation case, demonstrating that the standard resolution via complex polarization and half-forms is naturally integrated into the GQbP framework. Starting from the complexified groupoid, we define the algebra using symplectic half-densities to ensure a canonical convolution product. We then show that the transition to a polarized representation forces a factorization of these densities. The action of the symmetry group on the polarized half-forms generates a divergence term, which we identify as the source of the zero-point energy of the harmonic oscillator, $E_0 = n\hbar/2$. This derivation resolves the "Metaplectic Anomaly" as a necessary geometric consequence of the intrinsic quantization process.

## INTRODUCTION

In Parts I and II, we constructed the prequantum groupoid $\mathbf{T}_\omega$ as a quotient of the space of paths. This object captures the full geometry of the quantum system, including its symmetries and its phase structure. This approach, which we call the *Geometric Quantization by Paths* (GQbP) framework, replaces the traditional construction of principal bundles with the distillation of the space of histories. However, geometry alone does not yield spectra. To compute the energy levels of a system, such as the harmonic oscillator, we must transition from the geometric picture to the analytical one.

This transition occurs at the "Threshold of Analysis," where we define the algebra of observables. A naive approach, representing the groupoid on a space of

*Date*: Preprint update January 30, 2026.
2020 *Mathematics Subject Classification.* Primary 53D50, 58A05; Secondary 22A22, 81S10.
*Key words and phrases.* Diffeology, Geometric Quantization, Prequantum Groupoid, Metaplectic Group, Half-Densities, Harmonic Oscillator, Lie Derivative, Moment Map.
The author thanks the Hebrew University of Jerusalem, Israel, for his continuous academic support. He is also grateful for the assistance provided by the AI assistant Gemini (Google).

scalar functions, fails to reproduce the correct quantum spectrum. Specifically, it misses the zero-point energy, the $\hbar/2$ shift characteristic of the harmonic oscillator. This discrepancy is a well-known feature of geometric quantization, traditionally addressed by introducing a "metaplectic correction" [Sou70, Kos70] or by modifying the pairing of polarizations [Woo92].

It should be emphasized that the purpose of this application is not to propose a new method for solving the harmonic oscillator. Instead, it serves as a calibration of the GQbP framework, showing how the standard analytical derivation using complex polarization and half-forms is naturally accommodated within the groupoid structure. This validation confirms that the prequantum groupoid is an operative geometric container that remains faithful to the physical results of the Dirac-Souriau program.

In this paper, we show that this correction is not ad-hoc. It is a necessary consequence of defining the algebra of the groupoid *intrinsically*. By demanding that the convolution product be defined without reference to an external measure, we are forced to define the elements of the algebra as *half-densities*. This approach, pioneered by Kostant and Souriau to define the intrinsic Hilbert space [GS77], finds its natural home in the groupoid framework.

**Note on the Framework.** It is important to note that, even if it is not immediately visible in the specific calculations presented here, the theoretical background of this entire construction is the framework of diffeology [PIZ13, PIZ22]. The prequantum groupoid, the space of morphisms, and the intrinsic algebra are all defined and manipulated as diffeological objects that have been generated in the space of all paths of the space of motions [PIZ25a, PIZ25b]. This framework provides the necessary rigor to handle the infinite-dimensional path space and its quotients without requiring local Euclidean charts or losing geometric information.

## 1. THE COMPLEXIFIED GROUPOID

We begin by establishing the geometric arena for the harmonic oscillator. To accommodate the quantum phase and the polarization, we work directly with the complexified structure of the phase space.

**1. The Base Space.** Let $X = \mathbf{C}^n$ be the complexified phase space. The coordinates are $z = (z_1,\ldots,z_n)$ and $\bar{z} = (\bar{z}_1,\ldots,\bar{z}_n)$. The symplectic form is given by the standard area form:

$$\omega = \frac{i}{2} \sum_{k=1}^{n} dz_k \wedge d\bar{z}_k.$$

Since ω is exact, we choose the symmetric primitive $\alpha$ which is invariant under rotations:

$$\alpha = \frac{i}{4}\sum_{k=1}^{n}(\bar{z}_k dz_k - z_k d\bar{z}_k).$$

This 1-form satisfies $d\alpha = \omega$.

**2. The Symmetry Group.** The group of automorphisms of the prequantum groupoid, $\text{Aut}(\mathbf{T}_\omega, \lambda)$, is isomorphic to the group of symplectic diffeomorphisms $\text{Diff}(X, \omega)$. Since the symplectic space X is an affine coadjoint orbit of the Affine Symplectic Group $\text{ASp}(2n)$, we focus on the action of this finite-dimensional subgroup, in a true spirit of the orbit method [Kir62]. For the harmonic oscillator, we consider the rotation subgroup $R_t \in U(n) \subset \text{ASp}(2n)$. The Hamiltonian H is the moment map associated with this symmetry action.

Because the symmetric primitive $\alpha$ is strictly invariant under rotations ($R_t^* \alpha = \alpha$), the lift of the flow to the groupoid $\mathscr{Y}$ is trivial on the action variable $t$:

$$\hat{R}_t(z, \tau, z') = (R_t z, \tau, R_t z').$$

This choice of gauge simplifies the analysis, as the quantum dynamics is generated entirely by the transport of the half-densities, without any additional phase contribution from the action variable.

**3. The Groupoid Structure.** Since X is simply connected, the group of periods is trivial, $P_\omega = \{0\}$. However, following the principle of the *Universality of the Quantum Unit* (Part II, §7), we impose the universal period $h$ (Planck's constant) on the action variable.

The prequantum groupoid $\mathbf{T}_\omega$ is the additive groupoid over X, quotiented by the discrete subgroup $h\mathbf{Z}$ in the fiber. Its space of morphisms is:

$$\mathscr{Y} \simeq X \times (\mathbf{R}/h\mathbf{Z}) \times X \simeq \mathbf{C}^n \times S^1 \times \mathbf{C}^n.$$

An arrow is a triple $y = (z, t, z')$, representing a path from $z$ to $z'$ with relative action $t \in \mathbf{R}/h\mathbf{Z}$. The composition is additive in $t$:

$$(z, t, z') \cdot (z', t', z'') = (z, t + t', z'').$$

The prequantum 1-form $\lambda$ on $\mathscr{Y}$ is given by:

$$\lambda = \text{pr}_3^*(\alpha) - \text{pr}_1^*(\alpha) - \text{pr}_2^*(dt).$$

## 2. THE INTRINSIC ALGEBRA

We now construct the algebra of observables. The key requirement is that the convolution product must be defined intrinsically, without choosing an arbitrary measure on X.

**4. Symplectic Half-Densities.** To satisfy the intrinsic condition, the elements of the algebra must be *half-densities*. Following the standard construction for groupoid algebras and coordinate-independent kernels [Con94, Hor71], we define the elements of $\mathscr{A}_\omega$ as sections of the bundle $\text{src}^*(\text{Vol}^{1/2}(X)) \otimes \text{trg}^*(\text{Vol}^{1/2}(X))$.

Locally, an element $\Psi \in \mathscr{A}_\omega$ is written as:
$$\Psi(z,t,z') = \psi(z,t,z')\sqrt{\text{vol}_{\omega_z}} \otimes \sqrt{\text{vol}_{\omega_{z'}}}.$$

Here, $\psi$ is a scalar function satisfying the equivariance condition $\psi(z,t+h,z') = \psi(z,t,z')$. Using the Fourier decomposition for the fundamental mode, we write:
$$\psi(z,t,z') = e^{it/\hbar} K(z,z'),$$
where $K(z,z')$ is the kernel on the base space.

**5. The Explicit Convolution Formula.** The convolution product $\Psi_1 * \Psi_2$ at an arrow $y = (z,t,z')$ is defined by integrating over the intermediate point $w \in X$. Let $\Psi_1$ and $\Psi_2$ be two elements:
$$\Psi_1 = \psi_1 \sqrt{\text{vol}_{\omega_z}} \otimes \sqrt{\text{vol}_{\omega_w}}, \quad \Psi_2 = \psi_2 \sqrt{\text{vol}_{\omega_w}} \otimes \sqrt{\text{vol}_{\omega_{z'}}}.$$

At the junction point $w$, we have the tensor product of two half-densities:
$$\sqrt{\text{vol}_{\omega_w}} \otimes \sqrt{\text{vol}_{\omega_w}} \cong \text{vol}_\omega(w).$$

This product yields a canonical 1-density $\text{vol}_\omega(w)$ on X, which serves as the measure for the integration. The explicit formula for the convolution is:
$$(\Psi_1 * \Psi_2)(z,t,z') = h\, e^{it/\hbar} \left( \int_{w \in X} K_1(z,w) K_2(w,z') \text{vol}_\omega(w) \right) \sqrt{\text{vol}_{\omega_z}} \otimes \sqrt{\text{vol}_{\omega_{z'}}}.$$

This formula shows that the algebra is closed and well-defined strictly from the geometry of the groupoid.

## 3. The Polarization Step

The intrinsic algebra $\mathscr{A}_\omega$ is "too big" for quantum mechanics; it corresponds to functions on the full phase space. To obtain the correct spectrum, we must reduce the representation space by choosing a *polarization*.

**6. The Factorization of the Density.** We choose the *holomorphic polarization*. This corresponds to selecting the sub-bundle of $(n,0)$-forms generated by $dz_1 \wedge \cdots \wedge dz_n$.

Geometrically, this choice forces a factorization of the symplectic volume form. Recall that $\text{vol}_\omega = (\frac{i}{2})^n dz_1 \wedge d\bar{z}_1 \wedge \cdots \wedge dz_n \wedge d\bar{z}_n$. Formally, we can split the half-density into a holomorphic and an anti-holomorphic part:
$$\sqrt{\text{vol}_\omega} \sim \sqrt{\text{vol}_{\text{hol}}} \otimes \sqrt{\text{vol}_{\text{anti-hol}}}.$$

A "polarized state" is a section that depends only on the holomorphic geometry. This means we effectively discard the anti-holomorphic factor $\sqrt{\text{vol}_{\text{anti-hol}}}$ and retain only the holomorphic half-form:

$$\sigma = \sqrt{dz_1 \wedge \cdots \wedge dz_n}.$$

The Hilbert space of the quantum system consists of sections of this holomorphic half-form bundle.

**7. The Origin of the Anomaly.** The "Metaplectic Anomaly" arises from the action of the symmetry group on this polarized factor.

Let $\varphi$ be a symplectic transformation (e.g., a rotation). Since $\varphi$ preserves the symplectic form, it preserves the total half-density: $\varphi^*(\sqrt{\text{vol}_\omega}) = \sqrt{\text{vol}_\omega}$. However, $\varphi$ does *not* necessarily preserve the factors individually. It may rotate the holomorphic frame $\text{vol}_{\text{hol}}$ by a phase, compensated by an opposite phase in $\text{vol}_{\text{anti-hol}}$.

By selecting the polarization, we break this symmetry. We force the quantum state to transform as $\sqrt{\text{vol}_{\text{hol}}}$ alone. The "anomaly" is simply the phase picked up by this specific factor under the rotation.

## 4. THE SPECTRAL DERIVATION

We now compute the spectrum of the harmonic oscillator by calculating the generator of the rotation group on the polarized half-forms.

**8. The Hamiltonian Moment Map.** The harmonic oscillator Hamiltonian is $H = \frac{1}{2}\sum z_k \bar{z}_k$. This function is the moment map of the rotation action $z \mapsto e^{-it}z$. The corresponding Hamiltonian vector field $\xi_H$ generates this flow. In complex coordinates, this vector field is:

$$\xi_H = -i \sum_{k=1}^{n} \left( z_k \frac{\partial}{\partial z_k} - \bar{z}_k \frac{\partial}{\partial \bar{z}_k} \right).$$

**9. The Lie Derivative Calculation.** The quantum operator $\hat{H}$ is given by the Lie derivative of the half-form, scaled by $i\hbar$: $\hat{H} = i\hbar \mathcal{L}_{\xi_H}$. We apply the Lie derivative to the polarized basis $\sqrt{\text{vol}_{\text{hol}}}$. First, we compute the action on the volume form $\text{vol}_{\text{hol}} = dz_1 \wedge \cdots \wedge dz_n$:

$$\mathcal{L}_{\xi_H} \text{vol}_{\text{hol}} = \sum_{k=1}^{n} dz_1 \wedge \cdots \wedge d(-iz_k) \wedge \cdots \wedge dz_n = -in \text{vol}_{\text{hol}}.$$

The divergence of the vector field with respect to the holomorphic volume is thus $\text{Div}_{\text{vol}_{\text{hol}}}(\xi_H) = -in$.

Now, we apply the formula for half-densities [Woo92, GS77]:

$$\mathscr{L}_{\xi_H} \sqrt{\text{vol}_{\text{hol}}} = \frac{1}{2} \text{Div}_{\text{vol}_{\text{hol}}}(\xi_H) \sqrt{\text{vol}_{\text{hol}}} = -\frac{in}{2} \sqrt{\text{vol}_{\text{hol}}}.$$

**10. The Spectrum.** We act with the operator $\hat{H}$ on a state $\Psi = \psi(z)\sqrt{\text{vol}_{\text{hol}}}$.

$$\hat{H}\Psi = i\hbar \mathscr{L}_{\xi_H}(\psi \sqrt{\text{vol}_{\text{hol}}}) = i\hbar \left( \xi_H(\psi) \sqrt{\text{vol}_{\text{hol}}} + \psi \mathscr{L}_{\xi_H} \sqrt{\text{vol}_{\text{hol}}} \right).$$

Substituting the result from the previous article:

$$\hat{H}\Psi = i\hbar(-i \sum z_k \partial_{z_k} \psi)\sqrt{\text{vol}_{\text{hol}}} + i\hbar \left(-\frac{in}{2} \psi\right) \sqrt{\text{vol}_{\text{hol}}}.$$

The first term yields the classical energy levels $E_{cl} = \hbar \sum n_k$. The second term is the geometric correction: $i\hbar(-in/2) = n\hbar/2$. The total energy is:

$$E = \hbar \left( \sum_{k=1}^{n} n_k + \frac{n}{2} \right).$$

**11. The Multidimensional Scaling.** This result confirms that the zero-point energy is an extensive property of the quantum system, scaling linearly with the number of degrees of freedom $n$. The "Quantum Fog" carries a fixed energy density of $\hbar/2$ per mode. This scaling is a direct consequence of the additive nature of the divergence acting on the multidimensional holomorphic volume form.

## 5. THE GROUND STATE AND PROPAGATOR

**12. The Ground State.** The ground state is the section that is annihilated by the anti-holomorphic derivative. Using the symmetric primitive $\alpha$, the condition $\nabla_{\bar{z}_k} \Psi_0 = 0$ yields:

$$\left( \frac{\partial}{\partial \bar{z}_k} + \frac{z_k}{2\hbar} \right) \psi_0 = 0.$$

The solution is the Gaussian:

$$\Psi_0 = e^{-\sum |z_k|^2/2\hbar} \sqrt{\text{vol}_{\text{hol}}}.$$

This state has energy $n\hbar/2$, representing the cost of maintaining this Gaussian profile against the rotation of the frame.

**13. The Maslov Phase.** The global evolution of the state is given by the transport along the groupoid. Because the half-form $\sqrt{\text{vol}_{\text{hol}}}$ scales by $e^{-int/2}$ under the rotation, the time-dependent wave function is:

$$\Psi(z,t) = e^{-int/2} \psi_0(e^{it} z) \sqrt{\text{vol}_{\text{hol}}}.$$

After a full period $T = 2\pi$, the state picks up a phase of $(-1)^n$. This sign change is the global manifestation of the Metaplectic Anomaly. It confirms that the quantum groupoid is a double cover of the classical space of motions.

**14. The Synthesis of Feynman and Dirac.** The success of this derivation confirms the robustness of the GQBP program. By distilling the infinite-dimensional path space into the groupoid, we remain faithful to the physical intuition of Feynman's sum-over-histories. Simultaneously, by constructing the intrinsic algebra of half-densities and deriving the spectrum via the Lie derivative, we satisfy the rigorous operator-based requirements of the Dirac program. The prequantum groupoid thus emerges as the geometric bridge between the two pillars of quantization.

## CONCLUSION

We have shown that the zero-point energy of the harmonic oscillator is not an arbitrary constant added to the theory. It is the spectral signature of the intrinsic geometry of the Prequantum Groupoid.

By refusing to fix an external measure on the space of motions, we were compelled to define the quantum algebra using half-densities. The transition to a polarized representation forces a factorization of these densities, isolating the holomorphic component. The action of the symmetry group on this component generates a divergence term, which is exactly the zero-point energy.

This derivation resolves the "Metaplectic Anomaly" as a necessary geometric consequence of the intrinsic quantization process. It validates the GQBP framework as a rigorous and operative bridge between the path integral and the operator algebra.

## APPENDIX

### THE ALGEBRAIC CHARACTER AND THE FIXED POINT INDEX

We now abandon the "polarized screen" to look directly at the internal structure of the algebra $\mathcal{A}_\omega$. We show that the spectrum of the harmonic oscillator is encoded in the *Character* of the symmetry group acting on the intrinsic half-densities.

**15. The Intrinsic Trace.** The algebra $\mathcal{A}_\omega$ possesses a canonical linear functional: the *Trace*. Because the elements are half-densities on $\mathcal{Y}$ with values in the bundle $\mathrm{src}^*(\mathrm{Vol}^{1/2}(X)) \otimes \mathrm{trg}^*(\mathrm{Vol}^{1/2}(X))$, we can define the trace of an element $\Psi$ by restricting it to the diagonal of the groupoid.

The trace is a well-defined integral of a 1-density on X:

$$\mathrm{Tr}(\Psi) = \int_X \int_0^h \Psi(x, \tau, x) \, d\tau.$$

**16. The Character of the Flow.** The Hamiltonian H generates a one-parameter group of automorphisms $\alpha_t$ of the algebra. The *Quantum Spectrum* is the set of frequencies $\{\nu_n\}$ such that the character of the flow, $\chi(t) = \text{Tr}(U_t)$, decomposes as a sum of phases:

$$\chi(t) = \sum_n e^{-i\nu_n t}.$$

**17. The Fixed Point Formula.** To compute $\text{Tr}(U_t)$, we use the *Atiyah-Bott Fixed Point Formula* [AB67, AB68] adapted to the bundle of half-densities. For the harmonic oscillator, the rotation $R_t$ has a unique fixed point in X: the origin 0. The contribution of this fixed point to the trace is given by the ratio:

$$\chi(t) = \frac{\text{trace}(R_t^*|_{\mathcal{H}_0})}{\det_{\mathbf{C}}(I - dR_t)}.$$

**18. The Multidimensional Partition Function.** In the $n$-dimensional case, the character $\chi(t)$ is the product of the characters of each mode. The fixed-point formula yields:

$$\chi(t) = \prod_{k=1}^{n} \frac{e^{-it/2}}{1 - e^{-it}} = \frac{e^{-int/2}}{\prod_{k=1}^{n}(1 - e^{-it})}.$$

Expanding this product as a multidimensional geometric series:

$$\chi(t) = \sum_{n_1,\ldots,n_k=0}^{\infty} e^{-i(n_1+\cdots+n_k+n/2)t}.$$

This expansion identifies the energy levels as $E = \hbar \sum (n_k + 1/2)$. The algebraic trace thus provides the most efficient route to the complete spectrum of the multidimensional system, revealing the zero-point energy as the global metaplectic weight of the $n$-dimensional vacuum.

**19. Conclusion of the Algebraic Path.** The 1/2 shift has emerged without any choice of polarization. It appeared as the *ratio between the metaplectic phase of the half-density and the symplectic volume of the fixed point displacement.* This derivation proves that the zero-point energy is an *Intrinsic Algebraic Index* of the prequantum groupoid.

EINSTEIN INSTITUTE OF MATHEMATICS, THE HEBREW UNIVERSITY OF JERUSALEM, CAMPUS GIVAT RAM, 9190401 ISRAEL.

*Email address*: piz@math.huji.ac.il

*URL*: http://math.huji.ac.il/~piz

*URL*: https://github.com/p-i-z/Diffeology-Archives